\documentclass[a4paper,11pt,aip,jcp,amsmath,amssymb,reprint,nofootinbib,citeautoscript,showkeys]{revtex4-1}

\usepackage{extsizes}
\usepackage[left=.4in, right=.4in, top=1in, bottom=1in]{geometry}
\usepackage{balance}
\usepackage{times,mathptmx}
\usepackage{graphicx}
\usepackage{lastpage}
\usepackage[format=plain,justification=justified,singlelinecheck=false,font={stretch=1.125,small,sf},labelfont=bf,labelsep=space]{caption}
\usepackage{float}
\usepackage{fancyhdr}
\usepackage{fnpos}
\usepackage[english]{babel}
\addto{\captionsenglish}{}
\usepackage{array}
\usepackage{droidsans}
\usepackage{charter}
\usepackage[T1]{fontenc}
\usepackage[usenames,dvipsnames]{xcolor}
\usepackage{setspace}
\usepackage[compact]{titlesec}
\usepackage{hyperref}
\usepackage{epstopdf}
\definecolor{cream}{RGB}{222,217,201}
\usepackage{amsmath}
\usepackage{amssymb}
\usepackage{braket}

\DeclareMathOperator{\Tr}{Tr}
\DeclareMathOperator{\ci}{\mathrm{i}}
\DeclareMathOperator{\rmw}{\mathrm{w}}
\DeclareMathOperator{\rmd}{\mathrm{d}}
\DeclareMathOperator{\rmseo}{\mathrm{SEO}}

\newcommand{\mat}[1]{\ensuremath\mathsf{#1}}
\newcommand{\eu}[1]{\ensuremath\mathrm{e}^{#1}}

\newcommand{\Ketbra}[2]{\ensuremath\Ket{#1}\!\Bra{#2}}

\begin{document}

    \begin{abstract}
        The mapping approach addresses the mismatch between the continuous nuclear phase space
    	and discrete electronic states by creating an extended, fully continuous phase space
    	using a set of harmonic oscillators to encode the populations and coherences of the electronic states.
    	Existing quasiclassical dynamics methods based on mapping, such as the linearised semiclassical
    	initial value representation (LSC-IVR) and Poisson bracket mapping equation (PBME) approaches,
    	have been shown to fail in predicting
    	the correct relaxation of electronic-state populations following an initial excitation.
    	Here we generalise our recently published modification to the standard
    	quasiclassical approximation for simulating quantum correlation functions. We show that
    	the electronic-state population operator in any system can be exactly rewritten
    	as a sum of a traceless operator and the identity operator. 
    	We show that by treating the latter at a quantum level instead of using the mapping approach,
    	the accuracy of traditional quasiclassical dynamics methods
    	can be drastically improved, without changes to their underlying equations of
    	motion. We demonstrate this approach for the seven-state Frenkel-Exciton model
    	of the Fenna-Matthews-Olson light harvesting complex, showing that our
    	modification significantly improves the accuracy
    	of traditional mapping approaches
    	when compared to numerically exact quantum results.
    \end{abstract}
    
    \title{Improved population operators for multi-state nonadiabatic dynamics with the 
    mixed quantum-classical mapping approach}
    
    \author{Maximilian A.~C.~Saller}
    \affiliation{Laboratory of Physical Chemistry, ETH Zurich, Switzerland}
    
    \author{Aaron Kelly}
    \affiliation{Department of Chemistry, Dalhousie University, Halifax, Canada}
    
    \author{Jeremy O.~Richardson}
    \affiliation{Laboratory of Physical Chemistry, ETH Zurich, Switzerland}
    \email{jeremy.richardson@phys.chem.ethz.ch}
    
    \date{\today}
    
    \maketitle
    
    \section{Introduction}

	Simulating nonadiabatic effects in quantum dynamics continues to pose
	a considerable challenge in theoretical chemistry and physics, especially in
	the condensed phase. Arising when the energies of two or more electronic
	states approach each other, resulting in the breakdown of the Born--Oppenheimer
	approximation, these effects have been found to have a profound impact
	on a wide range of systems spanning physics, chemistry and biology.
	\cite{Marcus1993review, ChandlerET,Tully2012perspective}

	The development of simulation methods for nonadiabatic effects has thus
	continued to be the focus of considerable research efforts. Methods relying
	on an explicit expansion and propagation of the wavefunction, often on a
	grid, have yielded highly accurate results.
	\cite{Meyer1990MCTDH, MCTDH, Richings2015vMCG}
	However, many
	models inspired by condensed-phase systems still prove too computationally expensive to
	treat with these methods, due to their unfavourable exponential scaling with
	system size. Despite recent efforts to overcome this scaling hurdle,
	\cite{Martinez1998aims, Saller:2017:a}
	many systems from the fields of chemistry and
	biology, especially those in a condensed-phase environment, are simply too
	large to treat using a wavefunction-based approach. Mixed quantum-classical methods,
	\cite{Tully1990hopping,Sun1997mapping,Miller:1998:b,Stock2005nonadiabatic,Bonella2005LANDmap,Kapral:2008:a,Ananth2010mapping,Huo2011densitymatrix,Huo2012MolPhys,Kelly2012projectors,Hsieh2012FBTS,Hsieh2013FBTS,Ananth:2013:a,Richardson:2013:a,Richardson:2017:a,Agostini2014classical,Makri2015QCPI,Kapral2015QCL,Duke2015MVRPMD,Hele2016Faraday,Chowdhury2017CSRPMD}
	though inherently more approximate, are often the only choice when
	seeking to simulate nonadiabatic dynamics in the condensed phase.
	As many of these scale linearly with system size, they can readily be applied to
	large and complex realistic systems, yielding highly valuable insights
	at reasonable computational costs.

	The representation typically chosen for a nonadiabatic process consists of
	a continuous nuclear phase space and a set of discrete electronic states. The
	resulting Hamiltonian is given by
	\begin{equation}
		\hat{H} = \sum\limits_{j=1}^{F}\frac{\hat{p}_{j}^2}{2m_{j}} +
		U(\hat{\mat{x}}) + \hat{V}(\hat{\mat{x}})\,,
	\end{equation}
	where $\hat{p}_{j}$ and $m_{j}$ are the momentum operator and mass of nuclear
	degree of freedom (DoF) $j$ respectively, $\hat{\mat{x}}$ is a vector of
	length $F$ consisting of the position operators for each nuclear DoF.
	$U(\hat{\mat{x}})$ is the state-independent	potential and the state-dependent potential is
	given by
	\begin{equation}
		\hat{V}(\hat{\mat{x}}) = \sum\limits_{n,m}^{S} V_{nm}(\hat{\mat{x}})
		\Ketbra{n}{m}\,,
	\end{equation}
	where $S$ is the number of electronic states. The diagonal elements of
	$\hat{V}(\hat{\mat{x}})$ are the diabatic potential energy surfaces, while
	its off-diagonal elements are the couplings between the electronic states.
	Everything that follows does not rely on a particular choice of potential,
	\emph{i.e.}~we are not limited to simple harmonic models.
	Note that we will use reduced units throughout, such that $\hbar=1$.

	The mismatch between the continuous nuclear phase space and the discrete
	space of electronic states constitutes a recurring challenge in
	mixed quantum-classical dynamics. The mapping approach solves this problem by problem
	by projecting the electronic degrees of freedom into a space of
	singly-excited harmonic oscillators (SEOs).\cite{Meyer:1979:a,Stock1997mapping,Stock2005nonadiabatic}
	In the space of the SEO wavefunctions, the representation of state
	$\ket{n}$ is given by
	\begin{equation}\label{eq:mapwf}
		\braket{\mat{X}|n} = \frac{\sqrt{2}}{\pi^{S/4}} X_n \,
		\exp\left[-\frac{1}{2}\sum\limits_{m=1}^{S} X_{m}^2\right]\,,
	\end{equation}
	where $\mat{X}$ and its conjugate $\mat{P}$
	are vectors of length $S$, corresponding to the position and momenta
	of the SEOs.
	The mapping variables $\set{\mat{X}, \mat{P}}$
	extend the nuclear phase space, $\set{\mat{x}, \mat{p}}$. The resulting
	space, now completely continuous, can be used to propagate classical
	trajectories evolving under the mapping Hamiltonian, $\mathcal{H}$,
	given by
	\begin{equation}\label{eq:Hmap}
		\mathcal{H} = \sum\limits_{j=1}^{F}\frac{p_{j}^2}{2m_{j}}
		+ U(\mat{x}) + \frac{1}{2}\sum\limits_{n,m}^{S}
		\left( X_{n}X_{m} + P_{n}P_{m} - \delta_{nm} \right) V_{nm}(\mathsf{x}) \,.
	\end{equation}
	In addition to the relative simplicity of the
	mapping approach, the extended phase space grows linearly with the
	number of electronic states. Given furthermore the
	favourable scaling of classical trajectories with respect to the nuclear
	DoFs, a number of mixed quantum-classical dynamics approaches, aimed specifically at
	large, realistic systems in the condensed phase, have been developed based on
	this formalism.
	\cite{Sun1997mapping,Miller:1998:b,Stock2005nonadiabatic,Kapral:2008:a,Agostini2014classical,Kapral2015QCL}
	Note that in this work we will use the term quasiclassical to refer to
	mixed quantum-classical approaches which employ a single set of mapping
	variables per electronic state as well as a single set of
	positions and momenta for each nuclear degree of freedom.

	Quasiclassical methods yield accurate results for most observables at short
	times. In the long time limit however, they are well known to degrade in
	accuracy, especially for the relaxation to thermal equilibrium following
	an initial electronic excitation. Attempts to address this
	shortcoming with the use master equations have shown
	considerable promise.\cite{Markland:2015:a,Markland:2016:a}
    Other approaches to improve quasiclassical dynamics
    have led to the development of related dynamics approaches.
    For instance,
	the symmetrical quasiclassical windowing method
	uses a windowing function to ``bin'' the
	electronic populations, insuring that they have integer
	values at the beginning	and end of each trajectory.
	This approach has been applied to the benchmark we study
	below, achieving accuracy comparable to that reported here.
	\cite{Miller:2013:windowing,Cotton2013mapping,Miller2016Faraday,Miller:2016:d,Miller:2019:a}
    A number of methods which depart from the equations of motion underlying
    quasiclassical dynamics, but remain close to its overall motivation, have
    also shown considerable promise in treating multi-state systems.
    Prominent examples, which have been very successfully applied to the
    benchmark studied here, include the forward-backward trajectory solution
    \cite{Hsieh2012FBTS,Hsieh2013FBTS}
    (FBTS) and the partially linearised density matrix (PLDM) method.
    \cite{Huo2011densitymatrix,Huo2012MolPhys}
	
	In a recent publication, we have however shown that a simple modification,
	with a similar motivation as that underlying the use of master equations,
	can drastically improve the performance of quasiclassical methods, without
	changing the equations of motion.
	\cite{Saller:2019:a}
	We split the population operator into two parts, one of which is the
	identity.\cite{Kelly2012projectors}
	We can then use our understanding of the exact behaviour of this operator
	to drastically improve traditional quasiclassical methods.
	The resulting approach has the benefit
	of retaining all the advantages of these methods, as the underlying
	equations of motion are unchanged.

	Here we extend this approach, which was originally presented for only two
	electronic states, to an arbitrarily large electronic space.
	We use the fact that the electronic population, like any Hermitian operator, can be
	expanded exactly into the identity and a purely traceless component.
	\cite{Kelly2012projectors}
	Given that the behaviour of its quantum operator is well understood,
	we treat the identity exactly, resulting in a simpler phase-space representation
	of the population operator, involving only the traceless part.
	Population dynamics calculated using these
	modified operators are of drastically higher quality than those
	obtained from the traditional quasiclassical definition.

	We apply this general formulation to the challenging benchmark model
	for the Fenna-Matthews-Olson (FMO) light harvesting complex.
	\cite{Fenna:1975:FMO,Renger:2006:a,Ishizaki:2009:a,Ishizaki:2010:a}
	Our results are significantly more accurate than those obtained using the standard
	operator definitions and in excellent agreement with numerically exact
	quantum dynamics methods.

	\section{Theory}

	    Here we extend our previous work\cite{Saller:2019:a} by presenting a general formalism,
	    which can be applied to any system of multiple electronic states.

		\subsection{Quasiclassical population operators}
			In the mapping formalism, the operator $\Ketbra{n}{n}$, which measures the
			population of electronic state $\ket{n}$ can be written as
			\begin{equation}\label{eq:Pop}
				\Ketbra{n}{n} \equiv \hat{A}_n \mapsto \frac{1}{2}\bigg(\hat{X}_{n}^{2} +
				\hat{P}_{n}^{2} - 1\bigg)\,,
			\end{equation}
			where $X_{n}$ and $P_{n}$ are the mapping variables associated
			with state $\ket{n}$. In the quasiclassical approximation, the
			Wigner transform is used to define a phase-space representation for the
			operators of interest. The Wigner transform of a general operator,
			$\hat{O}$, is given by
			\begin{align}\label{eq:Wtrans}
				O^{\rmw} (\mat{x},\mat{p},\mat{X},\mat{P}) =
				&\iint \eu{\ci \mat{p}\cdot \mat{y} + \ci \mat{P}\cdot \mat{Y}}\nonumber\\
				&\Braket{ \mat{x}-\frac{\mat{y}}{2}, \mat{X}-\vspace{1em}\frac{\mat{Y}}{2} |
				\hat{O} | \mat{x}+\frac{\mat{y}}{2}, \mat{X}+\frac{\mat{Y}}{2} }
				\rmd \mat{y} \, \rmd \mat{Y}.
			\end{align}
			When considering the population operator, there are two representations
			one can choose to Wigner transform, corresponding to either the left
			or right-hand side of Eq.~\ref{eq:Pop}.
			Using the left-hand side is equivalent to including a projection on the SEO subspace,
			which yields an expression in terms of the harmonic oscillator wavefunctions as in
			Eq.~\ref{eq:mapwf}.\cite{Kelly2012projectors}
			The resulting phase-space representations, identified
			as $A_{n}^{\rmw}$ and $A_{n}^{\rmseo}$ respectively, are
			\begin{subequations}
				\begin{align}
					A_{n}^{\rmw}(\mat{X},\mat{P}) &= \frac{1}{2}
					\bigg(X_{n}^{2} + P_{n}^{2} - 1\bigg)\label{eq:Pw}\\
					A_{n}^{\rmseo}(\mat{X},\mat{P}) &= \frac{1}{2}
					\bigg(X_{n}^{2} + P_{n}^{2} -
					\frac{1}{2}\bigg)\phi(\mat{X},\mat{P})\label{eq:Pseo}\,,
				\end{align}
			\end{subequations}
			where
			\begin{equation}
				\phi(\mat{X},\mat{P}) = 2^{S+2} \exp\left[-\sum\limits_{m=1}^{S}
				(X_{m}^{2}+P_{m}^{2})\right]\,.
			\end{equation}
			Note that crucially, $A_{n}^{\rmw}\neq A_{n}^{\rmseo}\times\phi$.
			Each of these phase-space representations is derived \emph{via} the Wigner transform
			of a formally exact mapping form of the $\Ketbra{n}{n}$ operator.
		    Therefore, a clear choice of which to use when calculating
			observables is not obvious \textit{a priori}.
			A more detailed discussion of the possible combinations of phase-space
			representations and electronic initial conditions can be found
			in our recent work.
			\cite{Saller:2019:a}

			An observable commonly computed using quasiclassical methods is the
			population of a given electronic state, $\ket{n}$, given that the system
			was initially in a pure state, $\ket{m}$. In quantum mechanics this is
			defined by
			\begin{equation}
			    \label{eq:Pt}
				P_{n\leftarrow m}(t) = \Tr\bigg[\hat{\rho}_\text{b}\Ketbra{m}{m}
				\eu{\ci\hat{H}t}\Ketbra{n}{n}\eu{-\ci\hat{H}t}\bigg]\,,
			\end{equation}
			where $\hat{\rho}_\text{b}$ is a density matrix which defines the initial state of the nuclei,
			normalised such that the trace over nuclear DoFs only is $\Tr_{\mathrm{b}}[\hat{\rho}_{\mathrm{b}}]=1$.
			
		\subsection{Traceless projection operators}
		    
		    There are two differences between the phase-space representations given in
			Eq.~\ref{eq:Pw} and Eq.~\ref{eq:Pseo}:
			the factor of $\phi(\mat{X}, \mat{P})$, which is only present in $A_{n}^{\rmseo}$,
			and the differing constant terms, which are related to zero-point
			energy (ZPE) of the mapping DoFs.\cite{Mueller1999pyrazine}
			The origin of the latter is that both the projected and
			unprojected forms of $\Ketbra{n}{n}$ have a non-zero trace.
			We propose a form of the quantum population operator in which
			the trace is shifted to the identity operator,
			which in turn is treated exactly using quantum mechanics.
			\cite{Saller:2019:a}
			The result is a phase-space representation of the
			quantum population operator which is traceless.
			
			There is a unique expansion of the population operator $\Ketbra{n}{n}$,
			such that:
			\begin{equation}\label{eq:tr0pop}
				\Ketbra{n}{n} = \frac{1}{S}\bigg(\hat{\mathbb{I}} + \hat{Q}_{n}\bigg)\,,
			\end{equation}
			where $\hat{\mathbb{I}}=\sum_{m=1}^S\Ketbra{m}{m}$ is the identity operator,
		    and $\hat{Q}_{n}$ is,
			by design, traceless,
			\begin{equation}
			    \hat{Q}_n = 
				(S-1)\Ketbra{n}{n}-\sum\limits_{m\neq n}^{S}\Ketbra{m}{m}.
			\end{equation}
			Note that in a two-level system, this operator is the Pauli spin matrix,
			i.e. $\hat{Q}_1=\hat{\sigma}_z$,
			such that $\Ketbra{1}{1} = (\hat{\mathbb{I}}+\hat{\sigma}_{\mathrm{z}})/2$,
			which was used in our previous work.\cite{Saller:2019:a}
			Substituting this definition for the quantum population operator into
			Eq.~\ref{eq:Pt} and expanding yields
			\begin{align}\label{eq:bigPtr0}
				P_{n\leftarrow m}(t) =\frac{1}{S^2}\bigg( S &+
				\Tr\left[\hat{\rho}_{\mathrm{b}}\hat{\mathbb{I}}\,
				\eu{\ci\hat{H}t}\hat{Q}_{n}\eu{-\ci\hat{H}t}\right]\nonumber\\ &+
				\Tr\left[\hat{\rho}_{\mathrm{b}}\hat{Q}_{m}
				\eu{\ci\hat{H}t}\hat{Q}_{n}\eu{-\ci\hat{H}t}\right]\bigg)\,,
			\end{align}
			where we have used $\Tr[\hat{\rho}_{\mathrm{b}}\hat{Q}_{m}]=0$
			and $\Tr[\hat{\rho}_{\mathrm{b}}\hat{\mathbb{I}}]=S$.
            The final two terms in this expression are quantum correlation functions
            which can be approximated by well-known quasiclassical dynamics methods.

			Following the standard quasiclassical procedure, in order to calculate
			the value of the population operator given in Eq.~\ref{eq:bigPtr0},
			we Wigner transform the operators in these two constituent
			correlation functions.
			The phase-space representation of the traceless operator $\hat{Q}_{n}$ is

			\begin{equation}
				Q_{n}(\mat{X}, \mat{P}) =
				\frac{1}{2}\bigg[(S-1)(X_{n}^2 + P_{n}^2) -
				\sum\limits_{m\neq n}^{S}(X_{m}^2 + P_{m}^2)\bigg]\,.
			\end{equation}
			If we had performed the Wigner transform on the projected operator,
			the phase-space representation
            would simply be $Q_n(X,P)\phi(X,P)$.
            Note that either expression contains no constant terms
            which play the role of ZPE-parameters.

			It would be possible to arrive at a phase-space representation of the identity
			operator \emph{via} similar Wigner transforms. We however suggested in
			our previous work,\cite{Saller:2019:a} that we can instead use our
			understanding of its behaviour in quantum mechanics,
			which is to leave its operand unchanged.
			We therefore simply avoid directly computing the identity altogether.
			
			Starting from the exact expression for
			$\hat{P}_{n\leftarrow m}(t)$ in Eq.~\ref{eq:bigPtr0}, we thus arrive
			at our final quasiclassical expression for the population of electronic
			state $\ket{n}$, assuming the system was initially in state $\ket{m}$,
			\begin{equation}\label{eq:PfromQ}
				P_{n\leftarrow m}(t) \approx \frac{1}{S^2}\bigg(
				S + C_{\mathbb{I}Q_{n}}(t) + C_{Q_{m}Q_{n}}(t) \bigg)\,.
			\end{equation}
		    The constituent correlation functions $C_{\mathbb{I}Q_{n}}$ and
			$C_{Q_{m}Q_{n}}$ are given by
			\begin{subequations}\label{eq:newP}
    			\begin{align}
    			    C_{\mathbb{I}Q_n}(t) &= \big\langle\phi^a(\mat{X}, \mat{P})
    			    \, Q_n(\mat{X}(t),\mat{P}(t))\big\rangle\\
    			    C_{Q_mQ_n}(t) &= \Braket{\phi^a(\mat{X}, \mat{P})
    			    \, Q_m(\mat{X},\mat{P}) \, Q_n(\mat{X}(t),\mat{P}(t))}\,,
    			\end{align}
			\end{subequations}
			where $\braket{\cdots} = \frac{1}{(2\pi)^{F+S}}
			\iiiint \rho_b^{\rmw}(\mat{x},\mat{p}) \cdots \rmd\mat{x}\rmd\mat{p}\rmd\mat{X}\rmd\mat{P}$
			and $\rho_b^{\rmw}(\mat{x},\mat{p})$ is the Wigner transformed density matrix
			of the nuclear DOFs.
			In practice the values of these correlation functions are averaged
			over an ensemble of trajectories, with initial conditions for the 
			mapping variables being drawn from either $\phi(\mat{X}, \mat{P})$ or $\phi^{2}(\mat{X}, \mat{P})$, depending
			on whether the projected forms of one of both of $\hat{Q}_m$ and $\hat{Q}_n$ were
			Wigner transformed. This corresponds to $a=1$ and $a=2$ respectively.
			
			Note that we can include the factors of $\phi(\mat{X}, \mat{P})$ at
			time zero, because this function is constant over the course
			of any trajectory evolving under the Hamiltonian $\mathcal{H}$, given
			in Eq.~\ref{eq:Hmap}.
			Also the two constituent correlation functions can be calculated
			for all values of $m$ and $n$ in a single simulation.
			Just as in traditional quasiclassical methods,\cite{Miller:1998:b,Kapral:2008:a}
			the values of these constituent correlation
			functions, and therefore $P_{n\leftarrow m}(t)$, are exact in the limit
			of $t=0$.
			
		\subsection{Traditional quasiclassical dynamics methods}

            The traditional quasiclassical approach does not
            involve treating the identity quantum mechanically as we have 
            done above.
            There are two standard approaches which differ in
            whether both population operators are projected onto the subspace, or just one.
            These methods were derived in different ways \cite{Miller:1998:b, Kapral:2008:a}
            and are called
			the Poisson bracket mapping equation\cite{Kapral:2008:a,Kelly2012projectors} (PBME)
			and the linearized semiclassical initial value representation
			\cite{Sun1997mapping,Miller:1998:b} (LSC-IVR) methods.
			LSC-IVR commonly involves projecting both operators prior to Wigner
			transforming them, \emph{i.e.}~using $\Ketbra{m}{m} \mapsto A_{m}^{\rmseo}
			(\mat{X}, \mat{P})$ and $\Ketbra{n}{n} \mapsto A_{n}^{\rmseo}
			(\mat{X},\mat{P})$. The Wigner transform of each operator yields, as per
			Eq.~\ref{eq:Pseo}, a factor of $\phi(\mat{X}, \mat{P})$. Initial
			conditions for the mapping variables are therefore sampled from $\phi^2(\mat{X}, \mat{P})$.
			In PBME on the other hand, traditionally only the operator for the
			initial population is Wigner transformed in its projected from. The
			operators are therefore $\Ketbra{m}{m} \mapsto A_{m}^{\rmseo}
			(\mat{X}, \mat{P})$ and $\Ketbra{n}{n} \mapsto A_{n}^{\rmw}(\mat{X},\mat{P})$.
			Only the transform
			of $\Ketbra{m}{m}$ yields a factor of
			$\phi(\mat{X}, \mat{P})$. Consequently, electronic initial conditions
			are sampled from $\phi(\mat{X}, \mat{P})$. Using these definitions, the electronic population
			can be calculated from
    		\begin{subequations}\label{eq:tradP}
    			\begin{align}
    			    P^\text{PBME}_{n\leftarrow m}(t) &= \Braket{A_m^\text{SEO}(\mat{X},\mat{P}) \,
    			    A_n^\text{w}(\mat{X}(t),\mat{P}(t))}\label{eq:Ppbme}\\
    			    P^\text{LSCIVR}_{n\leftarrow m}(t) &= \Braket{A_m^\text{SEO}(\mat{X},\mat{P}) \,
    			    A_n^\text{SEO}(\mat{X}(t),\mat{P}(t))}\,.\label{eq:Plscivr}
    			\end{align}
    		\end{subequations}

			We note that
            the differences between these two methods does not actually stem from the derivations,
            but is mere convention.
            It would in principle be possible to derive a PBME method using two projections.            
			However for convenience, we will use Eqs.~\ref{eq:tradP} as the definition of PBME and LSC-IVR
			throughout this work.
			Finally, it is important to note that at least one of the operators has
			to be Wigner transformed in its projected form in order to
			ensure that the dynamics are initialised to the physical subspace.

			While both LSC-IVR and PBME, as well as other mixed quantum-classical
			methods, have been applied to challenging systems with considerable
			success, their failure to accurately reproduce population dynamics in the long
			time limit has been well documented. \cite{Mueller1998mapping,Bonella2003mapping,Kelly2012projectors,Saller:2019:a}
			As mentioned above, a number of modifications to quasiclassical methods
			which aim to address this issue have been proposed.
			\cite{Miller:2013:windowing,Cotton2013mapping,Miller2016Faraday,Miller:2016:d,Miller:2019:a,Cheng:2017:a,Saller:2019:a}

            In practice, both Eqs.~\ref{eq:tradP} and \ref{eq:newP} are evaluated
            by averaging over an ensemble of trajectories, propagated with Hamilton's
            equations of motion defined by $\mathcal{H}$. Initial conditions for
            each trajectory are sampled from $\rho_{\mathrm{b}}^{\rmw}(\mat{x}, \mat{p})$
            for the nuclei and $\phi^{a}(\mat{X}, \mat{P})$ for the mapping variables.

	\section{Results and Discussion}

		\subsection{The Fenna-Matthews-Olson Hamiltonian}

            The Fenna-Matthews-Olson complex is a pigment protein biomolecule
            found in green sulfur bacteria adapted for low-light environments.
            It consists of three identical trimers, each containing
            seven \emph{bacteriochlorophyll} (\emph{BChl}) pigments supported by a protein
            backbone. In photosynthesis the task of FMO is to transport the excitation
            gained from absorbing sunlight to the reaction centre where it is
            converted into electrochemical energy.
            \cite{Fenna:1975:FMO,Blankenship:1997:FMO,Renger:2006:a,Ishizaki:2009:a}

			The Frenkel-Exciton model for the energy transfer in FMO is a challenging
			benchmark for quantum dynamics methods.
			In comparison to the Spin-Boson systems studied in our previous work,\cite{Saller:2019:a}
			the FMO Hamiltonian presents a different kind of challenge to quasiclassical dynamics
			methods: the electronic subsystem is comprised of more electronic states and
			the system-bath coupling is different. We note that the key challenge resulting from
			a larger electronic state space is the possibility of reaction chains involving more
			than two states.
			As a result this benchmark and the FMO system in general has been extensively studied using a
			considerable number of approaches.
			\cite{Miller:2010:FMO,Huo2011densitymatrix,Kelly:2011:a,Kelly:2012:b,Hsieh2013FBTS,Miller2016Faraday,Miller:2016:d,Miller:2019:a,Cheng:2017:a,Scholes:2007:FMO,Adolphs:2007:FMO,Plenio:2008:FMO,Plenio:2010:FMO,OlayaCastro:2010:FMO,OlayaCastro:2011:FMO,Scholes:2011:FMO,Plenio:2013:FMO}
			In addition, though computationally challenging, numerically exact
			results are available, \emph{e.g.}~from hierarchical equations of motion (HEOM).
			\cite{Ishizaki:2009:a,Ishizaki:2010:a,Ishizaki:2010:b,FMO:HEOM:2011}
			
			In the seven-site model ($S=7$), the full FMO Hamiltonian is given by
			\begin{equation}\label{eq:FMOham}
				\hat{H}_{\mathrm{FMO}} = \hat{H}_{\mathrm{s}} + \hat{H}_{\mathrm{sb}} +
				\hat{H}_{\mathrm{b}}\,,
			\end{equation}
			where $\hat{H}_{\mathrm{s}}$ is the electronic sub-system Hamiltonian,
			given by
			\begin{equation}
				\hat{H}_{\mathrm{s}} = \sum\limits_{n=1}^{S} \varepsilon_{n}
				\Ketbra{n}{n} + \sum\limits_{n\neq m}^{S} \Delta_{nm}\Ketbra{n}{m}\,,
			\end{equation}
			where $\varepsilon_{n}$ is the energy of \emph{BChl} site $\ket{n}$ and
			$\Delta_{nm}$ is the electronic coupling between sites $\ket{n}$ and $\ket{m}$.
			The values of site energies and couplings used in the matrix
			representation of $\hat{H}_{\mathrm{s}}$ are given by
			\begin{equation}
				\!\!\!\!\!\!\!\!\!\!\!\!\hat{H}_{\mathrm{s}}\!=\!
				\left(\begin{array}{rrrrrrr}
					12410 & -87.7 &   5.5 &  -5.9 &   6.7 & -13.7 &  -9.9 \\
					-87.7 & 12530 &  30.8 &   8.2 &   0.7 &  11.8 &   4.3 \\
					  5.5 &  30.8 & 12210 & -53.5 &  -2.2 &  -9.6 &   6.0 \\
					 -5.9 &   8.2 & -53.5 & 12320 & -70.7 & -17.0 & -63.3 \\
					  6.7 &   0.7 &  -2.2 & -70.7 & 12480 &  81.1 &  -1.3 \\
					-13.7 &  11.8 &  -9.6 & -17.0 &  81.1 & 12630 &  39.7 \\
					 -9.9 &   4.3 &   6.0 & -63.3 &  -1.3 &  39.7 & 12440
				\end{array}\right)
			\end{equation}
			all energies being in units of cm$^{-1}$. The protein environment around
			every \emph{BChl} site is modelled by a bath of harmonic oscillators.
			The system-bath Hamiltonian, $\hat{H}_{\mathrm{sb}}$, defines the
			coupling between the electronic sub-system and these baths. It is given by
			\begin{equation}
			    \hat{H}_{\mathrm{sb}} = -\sum\limits_{n=1}^{S}\Ketbra{n}{n}
				\sum\limits_{j=1}^{B} c_{j}^{(n)} x_{j}^{(n)}\,,
			\end{equation}
			where $c_{j}^{(n)}$ is the vibronic coupling coefficient between site $\ket{n}$
			and bath mode $j$. $B$ is the number of modes per bath, such that $B = S\times F$.
			The position coordinate of bath mode $j$ of the $n$th bath is $x_{j}^{(n)}$.
			Finally, the Hamiltonian for the baths, $\hat{H}_{\mathrm{b}}$, is given by
			\begin{equation}
				\hat{H}_{\mathrm{b}} = \frac{1}{2}\sum\limits_{n=1}^{S} \Ketbra{n}{n}
				\sum\limits_{j=1}^{B} \left( \frac{\left(p^{(n)}_j\right)^{2}}{m_{j}} +
				m_{j}\left(\omega^{(n)}_{j}x^{(n)}_j\right)^{2}\right),
			\end{equation}
			where $p_{j}^{(n)}$ and $\omega^{(n)}_{j}$ are the momentum coordinate and frequency
			of bath mode $j$ associated with site $\ket{n}$.
			The choice of masses does not affect results, so one can effectively set $m_j=1$.
			Note that, following previous work,
			\cite{Ishizaki:2010:b,FMO:HEOM:2011,Kelly:2012:b,Miller:2019:a}
			each \emph{BChl} site is coupled to an identical bath, which
			in turn is uncoupled from all other baths. The coupling between
			sites is thus contained purely in $\hat{H}_{\mathrm{s}}$.

			The frequencies, $\omega^{(n)}_{j}$, and coupling coefficients, $c_{j}^{(n)}$,
			which are therefore identical for each bath,
			are drawn from a spectral density of the Debye form, given by
			\begin{equation}
				J(\omega) = 2\lambda\frac{\omega \omega_\text{c}}{\omega^2 + \omega_\text{c}^2}\,,
			\end{equation}
			where $\omega_\text{c}$ is the characteristic frequency of the bath, related
			inversely to the phonon relaxation time, $\omega_\text{c}^{-1} = \tau_\text{c}$,
			and we use $\lambda=35\,\mathrm{cm}^{-1}$ throughout, following previous
			work. \cite{Ishizaki:2010:b,FMO:HEOM:2011,Kelly:2012:b,Miller:2019:a}
			We discretize this function using a scheme known to reproduce exact
			reorganisation energies.\cite{Wang2001hybrid,Craig2007condensed}
			
			We define the initial nuclear density matrix $\rho_\text{b} = \eu{-\beta \hat{H}_\text{b}} / Z_\text{b}$,
			where the partition function $Z_\text{b}$ is defined such that the trace over
			bath modes only is $\Tr_\text{b}[\rho_\text{b}]=1$. Nuclear positions and momenta
			were sampled from the thermal Wigner distribution of the uncorrelated bath,
			given, for any bath, by
			\begin{align}
		        \rho_\text{b}^{\rmw}(\mat{x},\mat{p}) = \prod_{j=1}^{B}
		        &2\tanh\left(\tfrac{1}{2}\beta\omega_{j}\right)\nonumber\\
		        &\times\exp\left[-\tanh\left(\tfrac{1}{2}\beta\omega_{j}\right)
		        \left(\frac{p_{j}^{2}}{\omega_j}+\omega_{j}x_{j}^{2}\right)\right]\,.
			\end{align}

            \begin{figure*}[t!]
				\centering
				\includegraphics[width=6.74in]{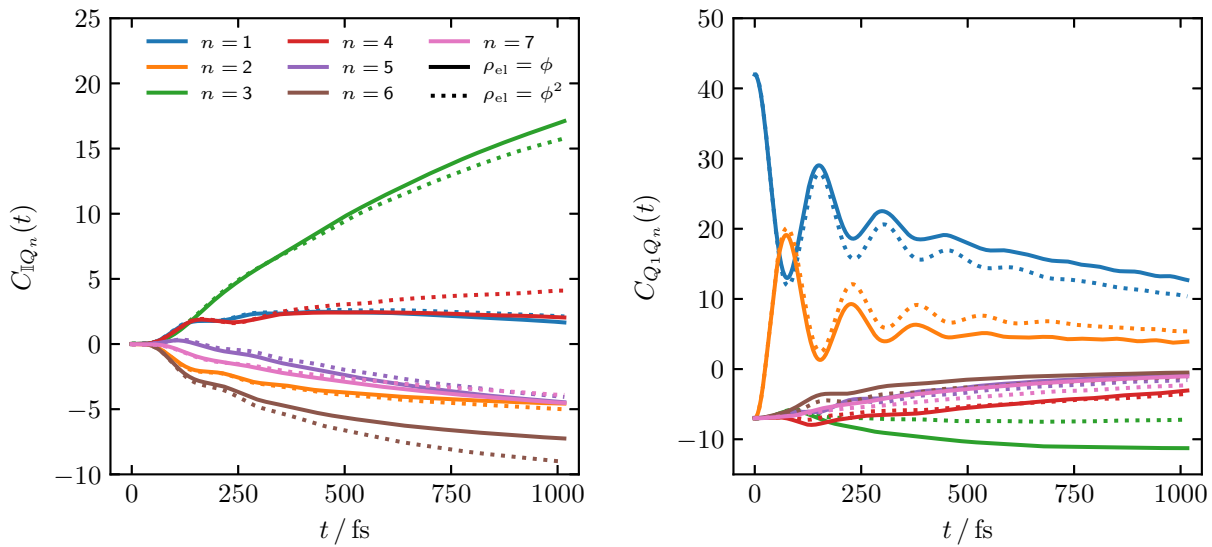}
				\caption{Constituent correlation functions of the FMO population, with
				$T=77\,\mathrm{K}$, and $\tau_{\mathrm{c}}=50\,\mathrm{fs}$ and an initial excitation of site $\ket{1}$.
				The solid and dotted lines correspond to the electronic initial
				conditions having been sampled either from $\phi(\mat{X},\mat{P})$
				or $\phi^{2}(\mat{X},\mat{P})$. In Eqs.~\ref{eq:newP}
				this corresponds to $a=1$ and $a=2$ respectively.
				\label{fig:constituents}}
			\end{figure*}

		\subsection{Simulation parameters}	
		
		    In order to test our alternative definition of the quasiclassical
            population in Eqs.~\ref{eq:newP}, we investigated three parameter
            regimes of the FMO Hamiltonian, which have been studied extensively,
            including using the numerically exact HEOM approach.
            \cite{Ishizaki:2009:a,Ishizaki:2010:a,Ishizaki:2010:b,FMO:HEOM:2011}
            All our simulations used a timestep of $\delta t=1 \text{fs}$, which
            was found to be numerically converged. The results
            presented here are averaged over an ensemble of $10^6$ trajectories
            in order to demonstrate the converged performance of our
            approach. We found however that using as few as $10^3$ trajectories
            was enough to qualitatively capture all significant features of
            the population dynamics and already exhibits the clear improvement
            over a fully converged traditional quasiclassical result.
            We note that in all our simulations, we used the traceless form
            of the $\hat{V}(\mathsf{x})$ matrix to propagate our trajectories,
            absorbing the remainder into $U(\mathsf{x})$.

		\subsection{Constituent Correlation Functions}
			
			Figure \ref{fig:constituents} shows the constituent correlation functions,
			$C_{\mathbb{I}Q_{n}}(t)$ and $C_{Q_m Q_n}(t)$, calculated with electronic
			initial conditions having been sampled from both $\phi(\mat{X}, \mat{P})$
			and $\phi^{2}(\mat{X}, \mat{P})$. 
			
			Considering the overall expression for the population
			given in Eq.~\ref{eq:PfromQ}, the magnitudes of the constituent
			correlation functions are as one might expect. Notably
			the negative values observed for both $C_{\mathbb{I}Q_{n}}(t)$
			and $C_{Q_m Q_n}(t)$ are not unphysical, as exact quantum
			mechanics would yield similar magnitudes for both correlation
			functions.
			We note that there is a noticeable difference
			between the correlation functions obtained from
			the two different initial distributions of the
			mapping variables we investigated.
			In order to assess their comparative accuracy however
			they must be combined, using Eq.~\ref{eq:PfromQ},
			into a population and compared to exact results.

		\subsection{Population dynamics}

			\begin{figure*}[t!]
				\centering
				\includegraphics[width=6.74in]{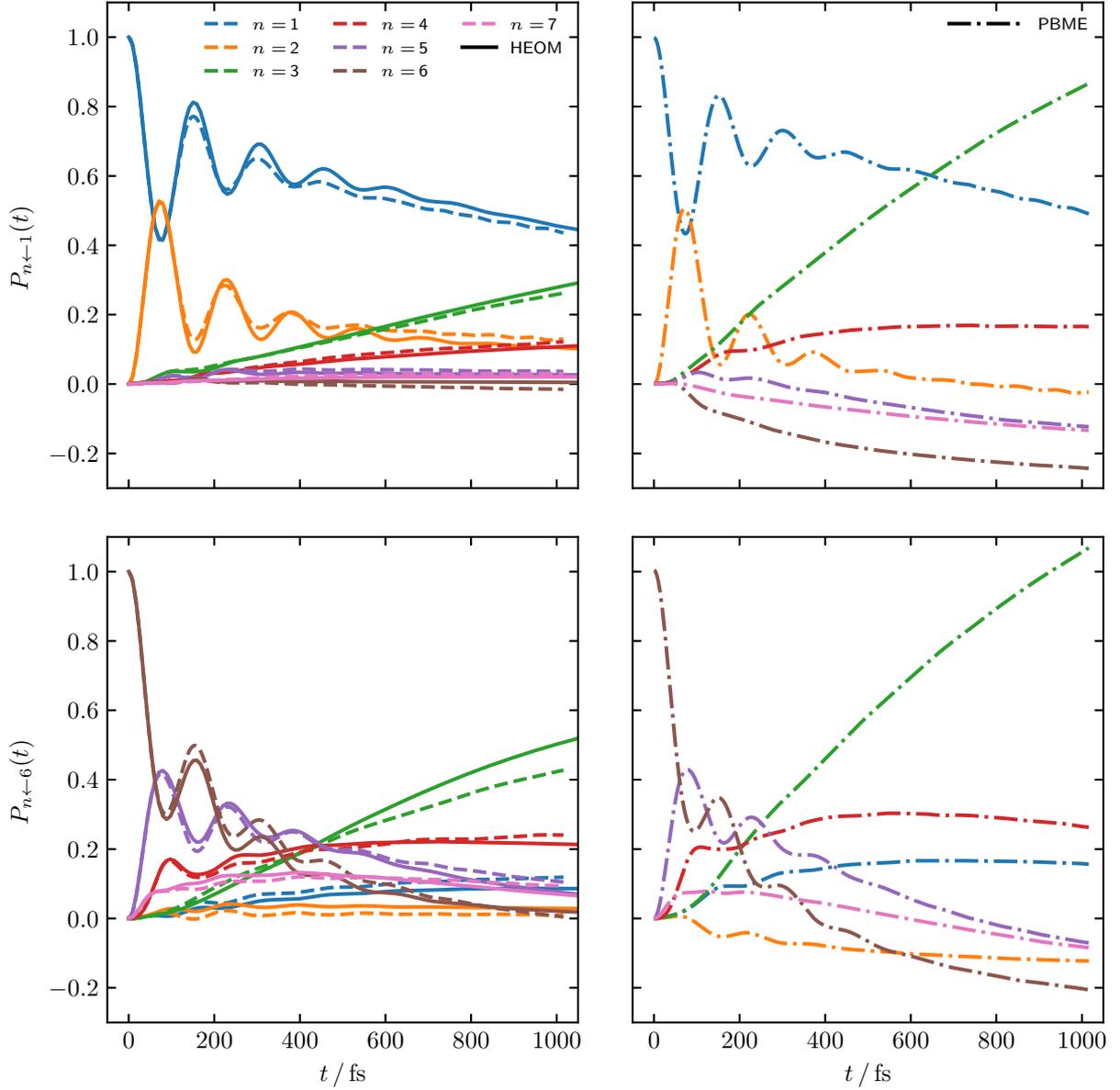}
				\caption{FMO site populations, with $T=77\,\mathrm{K}$, and
				$\tau_{\mathrm{c}}=50\,\mathrm{fs}$. Initial
				excitation of the $\ket{1}$ and $\ket{6}$ site are shown
				in the upper and lower panels respectively.
				Initial conditions for both
				constituent correlation functions sampled from $\phi(\mat{X}, \mat{P})$.
				Results using our alternative population operator are
				shown as dashed lines, traditional PBME are dash-dotted
				while solid lines are the numerically exact HEOM benchmark.
				\label{fig:pop_cold_single}}
			\end{figure*}

            Figure \ref{fig:pop_cold_single} shows the population dynamics
            resulting from combining the constituent correlation functions
            using initial conditions sampled from $\phi(\mat{X},\mat{P})$,
            \emph{i.e.}~the solid lines in Figure \ref{fig:constituents}.
            In addition to results obtained for an initial excitation of
            the $\ket{1}$ site, populations starting in site $\ket{6}$
            are also shown. In both cases the traditional PBME populations,
            calculated as per Eq.~\ref{eq:Ppbme}, are shown for comparison,
            along with numerically exact HEOM results.
            \cite{Ishizaki:2009:a,Ishizaki:2010:a,Ishizaki:2010:b,FMO:HEOM:2011}
            The comparison to PBME is a natural one here, as, in practice,
            the electronic initial conditions of PBME are also sampled
            from $\phi(\mat{X}, \mat{P})$, i.e.\ $a=1$. It is worth noting
            that this parameter regime of the FMO Hamiltonian is the most
			challenging for quasiclassical methods, due to the significant
			impact of quantum effects at low temperature.
            
            Our alternative definition of the population operators
            results in dynamics strikingly close in accuracy to the HEOM benchmark.
            We reproduce not only the correct ordering of states,
            even in the long time limit, but also capture all features
            present in the benchmark.
            We note that while we do observe negative populations
            for $P_{6\leftarrow 1}(t)$, their magnitude is almost negligible.
            In addition, our values for $P_{6\leftarrow 1}(t)$ are within
            the same margin of error of the exact HEOM result
            as every other population.
            This is especially encouraging when comparing the
            accuracy of our approach to that achieved with
            traditional PBME. The latter, while capturing 
            the population dynamics at short times rather well,
            completely fails to reproduce the long time behaviour.
            Notably the distribution and ordering of states
            beyond the short time limit degrades drastically with
            this approach.
            Considering that the low-temperature parameter regime of
            the FMO Hamiltonian poses a considerable challenge to
            quasiclassical methods, the accuracy of our results is
            highly encouraging.

			\begin{figure*}[t!]
				\centering
				\includegraphics[width=6.74in]{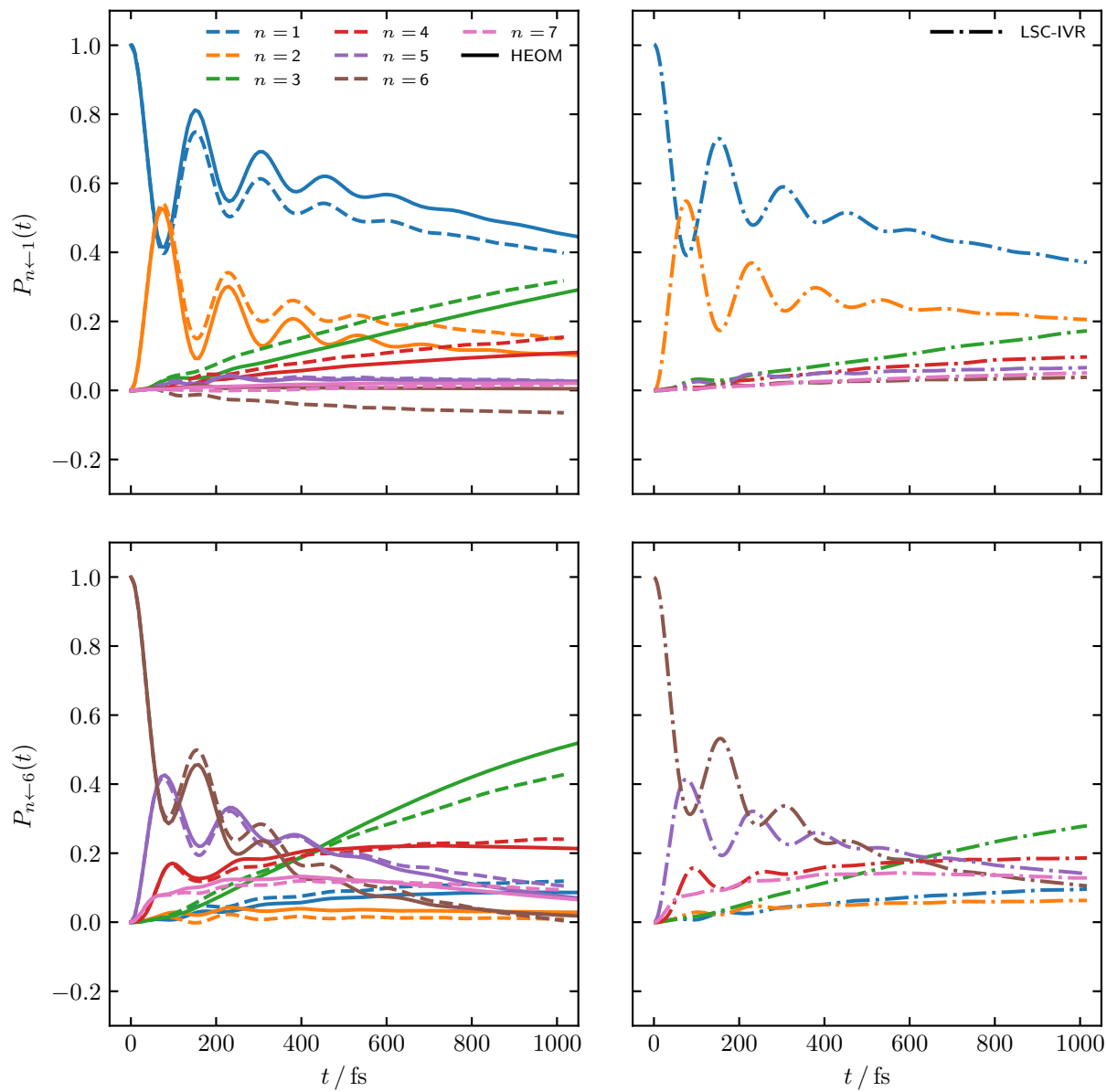}
				\caption{FMO site populations, with $T=77\,\mathrm{K}$, and
				$\tau_{\mathrm{c}}=50\,\mathrm{fs}$. Initial conditions for both
				constituent correlation functions sampled from $\phi^2(\mat{X}, \mat{P})$. As in
				Figure \ref{fig:pop_cold_single}, results using our
				approach are shown as dashed lines and the numerical HEOM
				benchmark as solid lines. The dash-dotted results
				are now the traditional LSC-IVR approach.
				\label{fig:pop_cold_double}}
			\end{figure*}

            Figure \ref{fig:pop_cold_double} shows populations for the same
            parameter regime of the FMO Hamiltonian as Figure \ref{fig:pop_cold_single}
            however with electronic initial conditions now having been sampled
            from $\phi^{2}(\mat{X}, \mat{P})$. This corresponds to adding
            the dotted lines of Figure \ref{fig:constituents} as per Eq.~\ref{eq:PfromQ}.
            Also shown are standard LSC-IVR results, which again are a natural
            source of comparison as they use the same electronic initial conditions.
            
            Comparing these results those shown in Figure \ref{fig:pop_cold_single},
            where the mapping variables were sampled from $\phi(\mat{X}, \mat{P})$,
            we observe a slight decrease in accuracy. Nevertheless, our approach
            retains all qualitative features of the dynamics and encouragingly
            yields the correct ordering of states throughout.
            LSC-IVR performs significantly better than PBME for this particular
            parameter regime, however still fails to yield accurate results
            beyond the short-time limit. We note in particular the
            incorrect ordering of states with respect to the HEOM
            benchmark.
            Our alternative definition of the population operator,
            although less accurate when sampling from $\phi^2(\mat{X}, \mat{P})$,
            therefore still improves considerably on the LSC-IVR result,
            which is highly encouraging.

			\begin{figure*}[t!]
				\centering
				\includegraphics[width=6.74in]{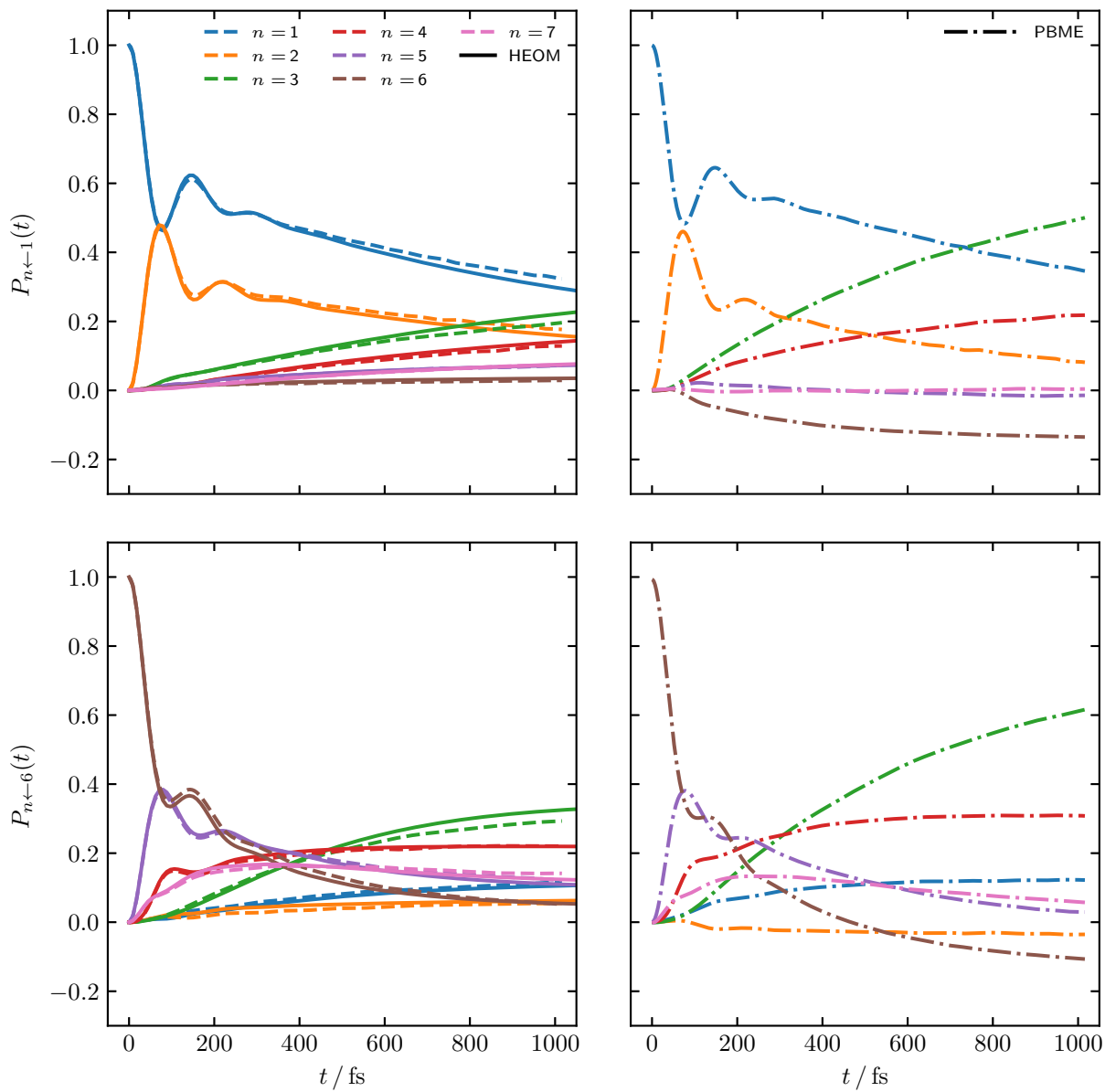}
				\caption{FMO site populations, with $T=300\,\mathrm{K}$, and
				a slower bath, $\tau_{\mathrm{c}}=50\,\mathrm{fs}$. Initial conditions
				for both constituent correlation functions sampled from $\phi(\mat{X}, \mat{P})$.
				Results are presented as in Figure \ref{fig:pop_cold_single}.
				\label{fig:pop_warm_slow}}
			\end{figure*}

			\begin{figure*}[t!]
				\centering
				\includegraphics[width=6.74in]{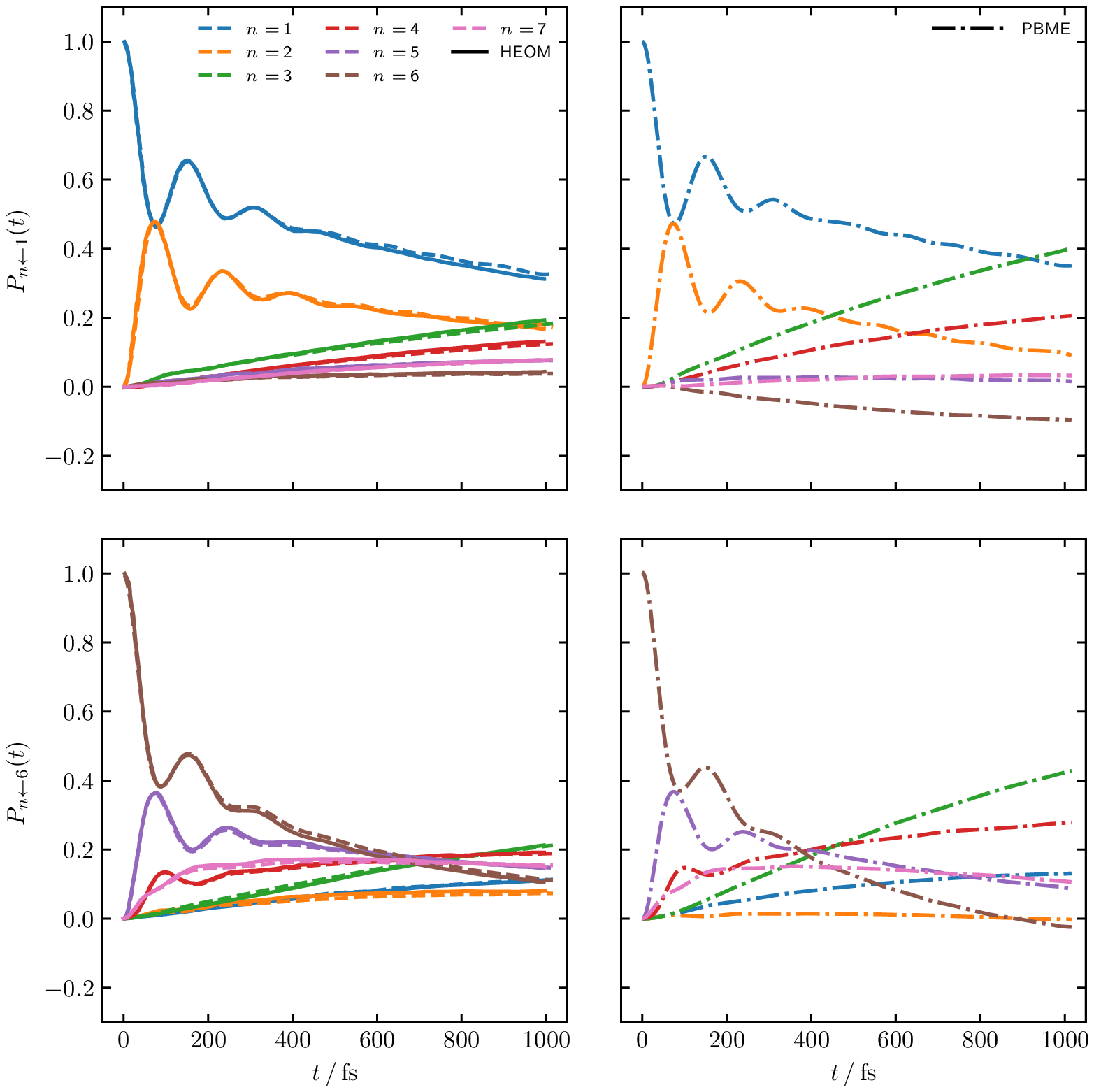}
				\caption{FMO site populations, with $T=300\,\mathrm{K}$, and
				a faster bath, $\tau_{\mathrm{c}}=166\,\mathrm{fs}$. Initial conditions
				for both constituent correlation functions sampled from $\phi(\mat{X}, \mat{P})$.
				\label{fig:pop_warm_fast}}
			\end{figure*}

            Figures \ref{fig:pop_warm_slow} and \ref{fig:pop_warm_fast} show
            the population dynamics of the FMO Hamiltonian, calculated with our
            alternative definition of the population operators and, as in
            Figure \ref{fig:pop_cold_single}, with electronic initial conditions
            sampled from $\phi(\mat{X},\mat{P})$, for two additional parameter
            regimes at $T=300$ K. Traditional PBME is again
            shown for comparison along with numerically exact HEOM
            benchmark results.
            \cite{Ishizaki:2009:a,Ishizaki:2010:a,Ishizaki:2010:b,FMO:HEOM:2011}
            We note that the dynamics in the parameter regimes shown in these two figures are,
            owing to the higher temperature, less likely to be affected by nuclear
            quantum effects.
            Nevertheless it is clear that in the long-time limit, PBME diverges significantly
            from the HEOM benchmark and yields an incorrect distribution of
            states.
            
            Using our alternative definition of the population operator
            again drastically improves the traditional quasiclassical result
            in both cases. Our approach in fact yields dynamics which
            now approach quantitative accuracy with respect to
            the exact HEOM benchmark. We furthermore note that
            the issue of small negative populations observed for our approach
            observed in Figure \ref{fig:pop_cold_single} has now
            disappeared. This is not surprising, given that low
            temperature systems are well known to constitute
            a more considerable challenge for quasiclassical
            approaches.
            We recognise that quasiclassical approaches are well
            known not to capture nuclear effects,
            owing to the fact that the trajectories underlying
            them are driven by classical equations of motion.
            We note however that our alternative definition 
            of the population operator is in fact not limited
            to quasiclassical methods, but may be applicable
            to other approaches based on mapping, which can
            capture nuclear quantum effects, such as nonadiabatic ring polymer
            molecular dynamics.
            \cite{Richardson:2013:a,Ananth:2013:a,Duke2015MVRPMD,Hele2016Faraday,Richardson:2017:a,Chowdhury2017CSRPMD}

		\subsection{Populations in the Long Time Limit}

			\begin{figure*}[t!]
				\centering
				\includegraphics[width=6.74in]{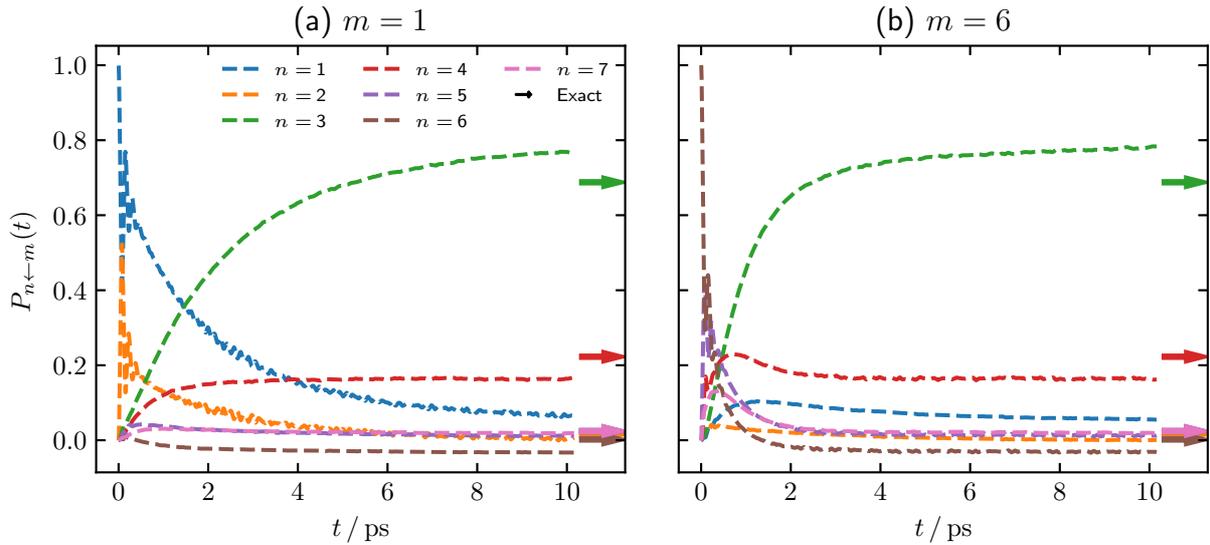}
				\caption{Long time FMO site populations, with $T=77\,\mathrm{K}$, and
				$\tau_{\mathrm{c}}=50\,\mathrm{fs}$. Initial conditions
				for both constituent correlation functions sampled from $\phi(\mat{X}, \mat{P})$.
				Long time limits were approximated using
				the diagonal elements of
				$\mathrm{e}^{-\beta\hat{H}_{\mathrm{s}}}/Z_\mathrm{s}$.
				\label{fig:pop_long}}
			\end{figure*}
			
			One of the well known failings of quasiclassical
			dynamics methods is that they do not preserve
			detailed balance of the populations and are
			therefore inaccurate at long times.	In order to
			investigate whether our alternative definition of
			the population operator can improve on this, we
			have carried out a longer simulation of the
			parameter regime shown in Figures \ref{fig:pop_cold_single}
			and \ref{fig:pop_cold_double}. We used the same
			timestep, $\delta t=1\,\text{fs}$, as in the
			simulations above and averaged over the same
			number of trajectories ($10^6$).
			Figure \ref{fig:pop_long} shows the FMO site populations,
			following an initial excitation of either the $\ket{1}$ 
			or the $\ket{6}$ site, up to 10 ps, calculated with
			our definition of the population operator. Electronic
			initial conditions for both were sampled from
			$\phi(\mat{X}, \mat{P})$.
									
			It is clear that our method does not rigorously preserve detailed balance
			as one of the populations is unphysically predicted to be slightly negative.
			We note however that as in Figures \ref{fig:pop_cold_single} and \ref{fig:pop_cold_double},
			our negative result is within the same margin of error of the exact result
			as every other state population.
			It is however well known that
			the long-time limits of the populations obtained
			from traditional quasiclassical approaches such as
			PBME and LSC-IVR can be much worse, predicting more strongly
			negative populations, and some greater than 1.
			\cite{Miller:2010:FMO}

            On the figure, arrows indicate an approximation to the distribution obtained 
			from the diagonal elements of the matrix exponential
			$\exp[-\beta \hat{H}_{\mathrm{s}}]/Z_\mathrm{s}$ where
			$Z_\mathrm{s}=\Tr[\exp(-\beta \hat{H}_{\mathrm{s}})]$.
			Note however that this is an approximation which neglects coupling to the bath modes.
			We are nevertheless encouraged that the equilibrium distribution predicted using our approach
			yields relatively similar results to this approximation.
			We note furthermore that our approach predicts identical equilibrium distributions,
			whether  site $\ket{1}$ or site $\ket{6}$ is initially excited.
			We have not computed
			the long-time dynamics of the other two parameter
			regimes we studied above. We do however expect
			that, owing to their higher temperature and thus
			the less impact nuclear quantum effects are
			likely to have in them, our approach would
			perform even better than in the $T=77$ K system.
			The fact that we do not observe negative populations
			in Figures \ref{fig:pop_warm_slow} and
			\ref{fig:pop_warm_fast} further supports this
			hypothesis.

		
		    Overall we consider these population results,
            along with the others shown above, to be highly
            encouraging. They clearly demonstrate
            that our definition of the population operator
            can drastically improve the accuracy of traditional
            quasiclassical approaches\cite{Miller:1998:b,Kapral:2008:a}
            at both intermediate and long times.

	\section{Conclusion}

        We have outlined an extension of our previous work,\cite{Saller:2019:a}
        presenting an alternative definition of the electronic population
        operator for any system of multiple electronic states.
        We rely on the fact that any Hermitian operator can be split
        into two terms, one of which is the identity, the other
        being traceless.\cite{Kelly2012projectors}
        We then use our understanding of the
        exact behaviour of the quantum identity instead of a
        quasiclassical treatment.
        The combination of this splitting and exact treatment
        of the identity results in a new form of the
        electronic population operator.
        Our approach retains the excellent scaling with
        respect to system size of traditional
        quasiclassical methods as well as their
        underlying equations of motion.
        Notably, as the constituent correlation functions
        into which our new operator is split
        can be calculated for all states in a single simulation,
        our approach is no more computationally expensive than
        the traditional methods it seeks to improve.
        
        We have applied our approach to the challenging seven-state
        Frenkel-Exciton model of the FMO light harvesting complex.
        \cite{Fenna:1975:FMO,Renger:2006:a,Ishizaki:2009:a,Ishizaki:2010:a}
        In addition to having been studied extensively with
        traditional quasiclassical methods,
        \cite{Miller:2010:FMO,Kelly:2011:a,Kelly:2012:b}
        the fact that numerically exact quantum results are available
        \cite{Ishizaki:2009:a,Ishizaki:2010:a,Ishizaki:2010:b,FMO:HEOM:2011}
        makes this system an ideal benchmark for our modification
        of the traditional quasiclassical population operators.
        
        Overall we find that using our alternative definition
        of the electronic population operator
        drastically improves on the results obtainable
        with other quasiclassical methods.
        In addition our results actually approach the exact
        quantum benchmark in accuracy for the three
        parameter regimes we study. Finally, we find that
        rather encouragingly, our method reproduces the
        long-time distribution of the electronic states with
        far higher accuracy than existing quasiclassical approaches.

        We recognise that there have been other efforts to fix the well documented
        shortcomings of traditional quasiclassical dynamics
        methods.
        For instance, the symmetrical quasiclassical windowing approach
        uses ``binning'' to convert the continuous mapping variables into integers,
        using a windowing function applied symmetrically at the
        beginning and end of each classical trajectory.
        \cite{Miller:2013:windowing,Cotton2013mapping,Miller2016Faraday,Miller:2016:d,Miller:2019:a}
        This approach has been applied to the FMO Hamiltonian we study here
        with considerable success, yielding results comparable in
        accuracy to those presented here.\cite{Miller2016Faraday,Miller:2016:d,Miller:2019:a}
        
        In recent work a post-processing method for the traditional LSC-IVR
        quasiclassical method has been proposed.\cite{Cheng:2017:a}
        The dynamics resulting from the traditional approach are shifted
        by a function which imposes the long-time Boltzmann distribution of
        the FMO subsystem Hamiltonian. We note that while the results
        obtained from this long-time correction do constitute 
        an improvement over the traditional approach, they
        fail to address any inaccuracies at short to medium times.
        In addition, this approach relies on either having prior
        knowledge of the correct long-time distribution of states
        or approximating it.

		Other mixed quantum-classical dynamics methods have also been applied to
		FMO Hamiltonian with considerable success. In recent work both the
		forward-backward trajectory solution\cite{Hsieh2012FBTS,Hsieh2013FBTS}
		and the partially linearised density matrix
		\cite{Huo2011densitymatrix,Huo2012MolPhys}
		approaches have yielded accurate results for the FMO
		systems studied here. We note however that these methods
		use a different set of equations of motion to propagate the
		classical trajectories from the quasiclassical approaches
		used here. This does however not disqualify them
		from also benefiting from our alternative definition of the
		population operator, as the latter is independent of
		the equations of motion.

        While we have shown that our alternative definition of the
        electronic population operator can drastically improve on the
        dynamics obtained from traditional quasiclassical methods,
        it cannot address all their shortcomings. Notably, it cannot capture nuclear quantum effects,
        which
        are fundamentally inaccessible to an approach relying
        on purely classical trajectories to calculate operators.
        Due to its simplicity and generality however, our approach
        could be combined with methods which can capture some nuclear
        quantum effects such as tunnelling.
        The nonadiabatic ring polymer molecular dynamics
        \cite{Richardson:2013:a,Ananth:2013:a,Duke2015MVRPMD,Hele2016Faraday,Richardson:2017:a,Chowdhury2017CSRPMD}
        method would seem to be a logical candidate for benefiting
        from our definition of the population operator, given
        that it is also based on
        the mapping formalism. We anticipate that the low
        temperature regime of the FMO Hamiltonian studied
        here may particularly benefit from such a combination,
        as nuclear quantum effects are likely to have a
        greater impact in this system.

        Overall we have shown that using our alternative definition of the population
        operator results in a considerable improvement over
        traditional quasiclassical approaches. Our results in
        fact approach the accuracy of the numerically exact benchmark for the
        systems we have studied. This is highly encouraging, given that
        the traditional methods we compare to have previously been
        considered inadequate for the simulation of long-time
        nonadiabatic dynamics. Our modification does not involve
        changing the equations of motion underlying quasiclassical
        methods and in fact also scales identically. We therefore hope that
        this work will further the development of dynamics methods
        based on the quasiclassical approach and of the
        progress of mixed quantum-classical dynamics as a whole.
        
	\section*{Acknowledgements}
	M.~A.~C.~S. would like to acknowledge financial support \textit{via} the
	ETH postdoctoral fellowship.  The authors also acknowledge the support from the
    Swiss National Science Foundation through the NCCR MUST (Molecular Ultrafast
    Science and Technology) Network.

	\bibliography{arxiv_ref}
	\bibliographystyle{rsc}

\end{document}